\documentclass[11pt,twoside]{article}

\usepackage{amsfonts,amsmath,amsthm,amssymb}
\usepackage{algorithm, algorithmic}

\usepackage{graphicx} 
\usepackage{subcaption}
\usepackage{caption}
\DeclareCaptionLabelFormat{figlabel}{Fig.~#2.}
\DeclareCaptionLabelFormat{tabformat}{Tab.~#2.}
\usepackage{enumerate}

\usepackage{diagbox} 
\usepackage{array}   

\usepackage{authblk}
\usepackage{color}
\usepackage{todonotes}
\usepackage{cancel}
\usepackage{url}
\usepackage{hyperref}
\usepackage{makeidx}
\usepackage{showidx}
\usepackage{multicol}        
\usepackage{xspace}
\usepackage{stmaryrd}        
\usepackage{pifont}          
\usepackage{fancybox}        
\usepackage{bm}

\theoremstyle{plain} 
\newtheorem{lem}{Lemma}[section] 
\newtheorem{defn}[lem]{Definition}
\newtheorem{thm}[lem]{Theorem}
\newtheorem{prop}[lem]{Proposition}
\newtheorem{cor}[lem]{Corollary}
\newtheorem{notn}[lem]{Notations}
\newtheorem{pb}[lem]{Problem}
\newtheorem{form}[lem]{Formulae}
\newtheorem{rk}[lem]{Remark}

\newtheorem*{com}{Comment}
\newtheorem*{ex}{Example}
\theoremstyle{remark}

\newcommand{\blem}{\begin{lem}}
	\newcommand{\elem}{\end{lem}}
\newcommand{\bdefn}{\begin{defn}}
	\newcommand{\edefn}{\end{defn}}
\newcommand{\bthm}{\begin{thm} }
	\newcommand{\ethm}{\end{thm}}
\newcommand{\bprop}{\begin{prop}}
	\newcommand{\eprop}{\end{prop}}
\newcommand{\bcor}{\begin{cor}}
	\newcommand{\ecor}{\end{cor}}
\newcommand{\bnotn}{\begin{notn}}
	\newcommand{\enotn}{\end{notn}}
\newcommand{\bpb}{\begin{pb}}
	\newcommand{\epb}{\end{pb}}
\newcommand{\bform}{\begin{form}}
	\newcommand{\eform}{\end{form}}
\newcommand{\brk}{\begin{rk}}
	\newcommand{\erk}{\end{rk}}
\newcommand{\bcom}{\begin{com}}
	\newcommand{\ecom}{\end{com}}
\newcommand{\bex}{\begin{ex}}
	\newcommand{\eex}{\end{ex}}
\newcommand{\bpf}{\begin{proof}}
	\newcommand{\epf}{\end{proof}}

\newcommand{\va}{{\bm a}}

\newcommand{\vd}{{\bm d}}
\newcommand{\ve}{{\bm e}}

\newcommand{\vg}{{\bm g}}
\newcommand{\vh}{{\bm h}}

\newcommand{\vu}{{\bm u}}
\newcommand{\vv}{{\bm v}}
\newcommand{\vw}{{\bm w}}
\newcommand{\vx}{{\bm x}}
\newcommand{\vy}{{\bm y}}
\newcommand{\vz}{{\bm z}}
\newcommand{\vA}{{\bm A}}
\newcommand{\vB}{{\bm B}}

\newcommand{\vI}{{\bm I}}

\newcommand{\cH}{\mathcal{H}}

\newcommand{\cL}{\mathcal{L}}

\newcommand{\cN}{\mathcal{N}}

\newcommand{\bR}{\mathbb{R}}

\newcommand{\be}{\begin{equation}}
	\newcommand{\ee}{\end{equation}}
\newcommand{\bmx}{\begin{matrix}}
	\newcommand{\emx}{\end{matrix}}
\newcommand{\bbmx}{\begin{bmatrix}}
	\newcommand{\ebmx}{\end{bmatrix}}
\newcommand{\bpmx}{\begin{pmatrix}}
	\newcommand{\epmx}{\end{pmatrix}}
\newcommand{\bvmx}{\begin{vmatrix}}
	\newcommand{\evmx}{\end{vmatrix}}

\newcommand{\ol}{\overline}

\newcommand{\ip}[2]{\left\langle {#1}, {#2} \right\rangle}
\newcommand{\nn}[1]{\left\| {#1} \right\|}
\newcommand{\nnn}[1]{\| {#1} \|_2}

\renewcommand{\t}[1]{{#1}^{\hspace{-0.2em}\top\!}} 

\newcommand{\supp}{\operatorname{supp}}

\newcommand{\argmin}{\mathop{\mathrm{argmin}}}

\renewcommand{\(}{\left(}
\renewcommand{\)}{\right)}

\newcommand{\al}{\alpha}
\newcommand{\la}{\lambda}

\newcommand{\eps}{\varepsilon}

\usepackage{mathtools}

\begin{document}

\title{Newton-Direction-Based ReLU-Thresholding Methods for Nonnegative Sparse Signal Recovery\thanks{This work was supported by the National Natural Science Foundation of China (12471295, 12426306, 12371305, 12301400),  Guangdong Basic and Applied Basic Research Foundation (2024A1515011566), Hetao Shenzhen-Hong Kong Science and Technology Innovation Cooperation Zone Project (HZQSWS-KCCYB-2024016), and Shandong Provincial Natural Science Foundation (ZR2023MA020, ZR2025MS99). }}

\author{Ning Bian\thanks{School of Mathematics and Statistics, Shandong University of Technology, Zibo, Shandong, China (\texttt{ningbian@sdut.edu.cn})},
 Zhong-Feng Sun\thanks{School of Mathematics and Statistics, Shandong University of Technology, Zibo, Shandong, China (\texttt{zfsun@sdut.edu.cn})},
 Yun-Bin Zhao\thanks{\emph{Corresponding author}. Shenzhen International Center for Industrial and Applied Mathematics, SRIBD, The Chinese University of Hong Kong, Shenzhen, China (\texttt{yunbinzhao@cuhk.edu.cn})}, Jin-Chuan Zhou \thanks{School of Mathematics and Statistics, Shandong University of Technology, Zibo, Shandong, China (\texttt{jinchuanzhou@sdut.edu.cn})}, 
 Nan Meng\thanks{Department of Mathematical Sciences, University of Nottingham Ningbo China, Ningbo, Zhejiang, China (\texttt{Nan.Meng@nottingham.edu.cn})}}

\maketitle

{\bf Abstract.}  

Nonnegative sparse signal recovery has been extensively studied due to its broad applications. Recent work has integrated rectified linear unit (ReLU) techniques to enhance existing recovery algorithms. We merge Newton-type thresholding with ReLU-based approaches to propose two algorithms: Newton-Direction-Based ReLU-Thresholding (NDRT) and its enhanced variant, Newton-Direction-Based ReLU-Thresholding Pursuit (NDRTP). Theoretical analysis iindicates that both algorithms can guarantee exact recovery of nonnegative sparse signals when the measurement matrix satisfies a certain condition.. Numerical experiments demonstrate NDRTP achieves competitive performance compared to several existing methods in both noisy and noiseless scenarios.



\section{Introduction} 

A fundamental task of compressive sensing is to recover (or reconstruct) the sparse signal from limited  measurements. 
In many situations, the signals are not only sparse but also nonnegative. In such cases, one need to handle the nonnegative sparse signal recovery, which has been widely  studied in various areas such as geoscience and remote sensing \cite{MJA, EYJ},
medical imaging \cite{PS} and face recognition \cite{ JLZ,  NBS, RWBX, DUSYS}.

Given a measurement matrix $\vA \in \bR^{m \times n}$ with $m \ll n$, the inaccurate measurements with noise $\ve \in \bR^m$ of a nonnegative sparse signal 
$\vx \in \bR^n$  are obtained as
\be \label{model}
\vy  = \vA \vx + \ve.
\ee
The nonnegative sparse-signal-recovery problem is concerning about the recovery of  $\vx \in \bR^n$  from the acquired measurements $\vy \in \bR^m$ which are significantly smaller than the signal length. This problem can be  formulated as the following sparse  optimization problem with a nonnegative constraint:
\be  \label{model1}
\min_{\vx \in \mathbb{R}^n} \left\{\frac{1}{2} \nnn{\vy - \vA \vx}^2 \  : \  
 \nn{\vx}_0 \le k, \vx \ge \bf{0} \right\},
\ee
where  $\nn{\cdot}_0$ is the '$\ell_0$-norm' counting the number of nonzero entries.
It has been shown in \cite{SD,Zh2} that  the  exact recovery of nonnegative sparse signals can be achieved through solving an $\ell_1$-regularized nonnegative least squares (NNLS) problem. The classical active-set algorithm proposed by Lawson and Hanson \cite{CR} is one of the efficient methods for such optimization problems. Their algorithm for NNLS has become a fundamental tool in nonnegative sparse recovery due to its guaranteed convergence to an optimal solution to the problem.
Various algorithms based on NNLS have been introduced to solve (\ref{model}), such as hard thresholding NNLS \cite{MM, MM2}, sparse NNLS \cite{RF}, and nonnegative orthogonal least squares \cite{TJCE}. 
Nonnegative orthogonal matching pursuit (NNOMP) \cite{AEM} refines the selected coefficients by solving  NNLS subproblems. Later, a fast implementation of NNOMP called fast NNOMP \cite{MDM} was proposed. 
Several nonnegative variants of the standard greedy sparse-signal-recovery algorithms have been introduced in \cite{KH}, including compressive sampling matching pursuit, subspace pursuit, and hard thresholding pursuit.  
For convenience of discussion in this paper, we refer to the nonnegative variant of subspace pursuit as Nonnegative Subspace Pursuit (NNSP).
It has been shown that nonnegative sparse signals can be exactly recovered via $\ell_1$-minimization if the null space of the measurement matrix satisfies a certain condition \cite{KDXH, Zhang}. In addition,  Zhao \cite{Zh2,Z18}  introduced the Range Space Property of transposed matrix to characterize the uniform recovery of nonnegative sparse signals, offering a complementary perspective to traditional null space analysis.

The rectified linear unit (ReLU) function was first introduced by  Fukushima's neocognitron model \cite{FK}, inspired by biological neurons' activation thresholds. It gained prominence in 2010s with deep learning breakthroughs, replacing Sigmoid/Tanh due to its sparsity induction and vanishing gradient mitigation \cite{VG}. 
ReLU enforces non-negativity and sparsity by zeroing negative inputs, making it ideal for recovering non-negative sparse signals. Its computational efficiency and gradient stability accelerate training and improve convergence in deep networks for signal reconstruction.
Recently, some algorithms combining the ReLU function with hard thresholding have been proposed for nonnegative signal recovery, such as the ReLU-based hard thresholding (RHT) \cite{HSWW}, momentum-boosted adaptive thresholding \cite{ZZJH} and ReLU-based hard thresholding pursuit (RHTP)  \cite{PSRP}.

The Newton-type method is a powerful numerical method for optimization problems, which exhibits local quadratic convergence and often achieves  high-accuracy solutions using far fewer iterations compared to the first-order methods.  Recently, it has been extended to solve the sparse-signal-recovery problem (see, e.g., \cite{NM}). The basic idea to extend the classic Newton method to sparse signal recovery is to apply a certain thresholding such as the hard thresholding to an iterate generated by the Newton search direction. Specifically, they take the following iterative scheme:
\be \label{NSHT_iter}
\vx^{(p+1)} = \cH_k \(\vx^{(p)} + \la (\t{\vA} \vA + \eps \vI)^{-1} \t{\vA}(\vy - \vA \vx^{(p)})\),
\ee   
where $\vx^{(p)}$ is the current iterate, $\vx^{(p+1)}$ is the next iterate, $\la>0$ is the stepsize, $\eps>0$  is a regularization parameter introduced to avoid singularity of the matrix, and $\cH_k(\cdot)$ is the hard thresholding operator which keeps the $k$ largest-magnitude entries of a vector and zeros its  remaining entries.
Following this scheme, \cite{NM} presents two novel algorithms: Newton-Step-Based Iterative Hard Thresholding (NSIHT) and Newton-Step-Based Hard Thresholding Pursuit (NSHTP), which are empirically demonstrated to be effective and stable. 
However, their convergence analysis requires $\eps > \sigma_1^2$, where $\sigma_1$ is the largest singular value of the measurement matrix $\vA$. When $\sigma_1$ is large, this constraint obliges $\eps$ to be large as well,  so $\t{\vA}\vA + \eps \vI$ can not be regarded as a small perturbation of the true Hessian $\t{\vA}\vA$ of the objective function in (\ref{model1}).
This may compromise the advantages of Newton's method by preventing it from fully leveraging the Hessian information.

For nonnegative sparse signal recovery, we propose the so-called Newton-Direction-Based ReLU-Thresholding (NDRT) algorithm, which is a modification of the framework in (\ref{NSHT_iter}) through merging the ReLU operation,  Newton step and  hard thresholding. By further incorporating a nonnegative projection step, we develop the so-called Newton-Direction-Based ReLU-Thresholding Pursuit (NDRTP). Furthermore, our convergence analysis eliminates the restrictive condition $\eps > \sigma_1^2$ used in \cite{NM} , and hence our method allows $\eps$ to be an arbitrarily small positive number. The main contribution of the paper includes:
\begin{itemize}
  \item By integrating the Newton direction, ReLU thresholding and hard thresholding, we propose NDRT and its enhanced variant NDRTP for the nonnegative sparse signal recovery problem (\ref{model1});  see Section 2 for details.

  \item  Under certain assumptions on the measurement matrix, we develop sufficient conditions to ensure the guranteed performance  of the proposed algorithms in nonnegative sparse signal recovery. The main theoretical guarantees are presented in Theorems~\ref{convofNDRT} and~\ref{convofRNHTP}. 
      Our spectral-decomposition-based analysis allows the algorithmic parameter $\eps$ to be small, so that $\t{\vA}\vA + \eps \vI$ a truly slight perturbation of the Hessian $\t{\vA}\vA$.

  \item Leveraging the Karush–Kuhn–Tucker (KKT) conditions, we establish a fundamental error bound associated with nonnegative projection problem, summarized in Theorem~\ref{KeyLem}, which plays a crucial role in establishing the guaranteed performance of NDRTP.

   \item Through simulations, we compare the performance of the proposed algorithms with several existing algorithms in  nonnegative-sparse-signal recovery. Moreover, we propose a very helpful empirical formula for the iterative stepsize $\la$ in  NDRTP. 
\end{itemize}

The remainder of this paper is organized as follows. Section 2 introduces some fundamental inequalities and outlines our algorithms. Section 3 performs a theoretical analysis of the algorithms. Section 4 presents some numerical results, and Section 5 concludes the paper.

\section{Preliminaries and Algorithms}

This section introduces the notation, reviews several fundamental inequalities, and then presents the proposed algorithms for nonnegative sparse signal recovery.

\subsection{Notation and basic inequalities}

We begin by introducing some notation used in the paper.
Let $[n]$ denote the index set $\{1, 2, \cdots, n\}$ for any positive integer $n$. For a given subset $\Omega$ of $[n]$, 
the complement of $\Omega$ in $[n]$  is written as $\ol{\Omega} := [n] \setminus \Omega$ and  $|\Omega|$ represents
 the cardinality of $\Omega$.  Let $\vx := \t{(x_1, \dots, x_n)}   \in \bR^n$ be a vector.
 We call $\vx$ is $k$-sparse if $\nn{\vx}_0 \le k$, where  $\nn{\cdot}_0$ is the '$\ell_0$-norm' counting the number of nonzero entries.
The support of $\vx$ is denoted by $\supp(\vx)$, defined as $\supp(\vx) := \{i \in [n] : x_i \neq 0\}$.
The index set of the $k$ largest  entries in magnitude of $\vx$ is denoted by $\cL_k(\vx)$. We denote the positive part of $\vx$ by $\vx^+$, where $\vx^+ : = \t{(\max\{x_1, 0\}, \dots, \max\{x_n, 0\})}$ and  denote its negative part by
$\vx^- : = \vx - \vx^+$. 
The ReLU function  $\Psi(\vx)$ is defined as $\Psi(\vx):=\vx^+$.
For any index set $\Omega \subseteq [n]$, $\vx_\Omega$ refers to the vector obtained by retaining entries indexed by $\Omega$ and zeroing out the others. 
The $\ell_2$-norm $\nnn{\vx}$ is defined as $\nnn{\vx} := \sqrt{\t{\vx} \vx}$. The spectral norm of matrix $\vA$ 
is written as $\nnn{\vA}$, defined as its largest singular value. 
 Given a real number $t$,  $\lceil t \rceil$ denotes the smallest integer not less than $t$. 
For convenience, we denote the golden ratio by $\phi$,  i.e., $\phi =(\sqrt{5}+1)/2 \approx 1.618$.

We first recall some fundamental concepts and inequalities that are very useful in our later analysis.

\bdefn [See \cite{CT}]
Let $\vA \in \bR^{m \times n}$ be a matrix with $m \ll n$. The restricted isometry constant (RIC) of order $k$, denoted
by $\delta_k$, is the smallest number $\delta \ge 0$ such that
$$(1-\delta) \nnn{\vx}^2 \le \nnn{\vA \vx} \le (1 + \delta) \nnn{\vx}^2$$ 
for all $k$-sparse vectors $\vx \in \bR^n$. If $\delta_k < 1$, then $\vA$ is said to satisfy the restricted isometry property (RIP) of order $k$.
\edefn

\blem [See \cite{SF}] 
Let $\vA \in \bR^{m \times n}$ with $m \ll n$ be a matrix. Let $\vu \in \bR^m$, $\vv, \vw \in \bR^n$, and let $k$ be a positive integer. Then the following inequalities hold:
\begin{flalign}
{\rm (a)}\ & \text{For any index set}~\Omega \subseteq [n]~\text{with}~|\Omega|\le k, 
 \ \nnn{( \t{\vA} \vu )_\Omega} \le \sqrt{1+\delta_k}\, \nnn{\vu}. & \label{RIC_ineq1}\\
{\rm (b)}\ & \text{For any index set}~\Gamma \subseteq [n]~\text{with}~|\Gamma\cup \supp(\vv)|\le k, 
 \ \nnn{((\vI - \t{\vA} \vA)\vv )_\Gamma} \le \delta_k\, \nnn{\vv}. & \label{RIC_ineq2}\\
{\rm (c)}\ & \text{If}~|\supp(\vv)\cup \supp(\vw)|\le k, ~\text{then}~  
| \langle {\vv}, (\vI-\t{\vA}\vA)\vw \rangle | \le \delta_k\, \nnn{\vv}\, \nnn{\vw}. & \label{RIC_ineq3}
\end{flalign}
\elem


\blem [See \cite{ZH}] \label{goldenration}
For any  vector $\vu\in \bR^n$ and $k$-sparse vector $\vx \in \bR^n$, we have
\be \label{tightbound}
\nnn{ \cH_k(\vu)- \vx} \le \phi  \nnn{(\vu - \vx)_\Omega},
\ee
where $\Omega := \supp(\cH_k(\vu)) \cup \supp(\vx)$ and $\phi =(\sqrt{5}+1)/2$ (the golden ratio).
\elem

\subsection{Algorithms}

The Hessian matrix of the objective function in (\ref{model1}) is $\t{\vA}\vA$,  
which is singular because $\vA\in \bR^{m \times n}$ with $m\ll n$.
To address this issue, the Newton-step-based hard-thresholding algorithms proposed in \cite{NM} employ a modified Hessian $\t{\vA} \vA + \eps \vI$ with $\eps > 0$.  This modification yields the iterative scheme (\ref{NSHT_iter}) for sparse optimization problems. 
However, the iterates $\vx^{(p+1)}$ generated by this scheme may violate nonnegativity constraints, and thus this method cannot directly apply to solve the problem (\ref{model1}).   To enforce nonnegativity, we introduce a ReLU step before performing hard thresholding in each iteration. In other words, we projects the resulting vector produced by the modified Newton method onto the set of nonnegative $k$-sparse vectors. This is achieved by retaining the largest $k$ nonnegative entries  and setting all other entries to zero, thereby yielding a $k$-sparse nonnegative vector at every step.
Thus we obtain the following iterative scheme: 
\be \label{RNS_iter}
\vx^{(p+1)} = \cH_k \( \Psi \(\vx^{(p)}  + \la (\t{\vA} \vA + \eps \vI)^{-1} \t{\vA}(\vy - \vA \vx^{(p)})\) \),
\ee
where $\Psi(\cdot)$ is the ReLU function.
A direct implementation of this scheme leads to the NDRT algorithm,  
which is summarized in \emph{Algorithm \ref{RIHTwithnewtonstep}}. Its enhanced version obtained by incorporating an additional nonnegative projection operator at each iteration is called NDRTP, which is outlined in \emph{Algorithm \ref{RHTPwithnewtonstep}}. In NDRTP, step (\ref{NNLS})  solves an NNLS problem, which is handled by the gradient projection method in practice.

\begin{algorithm} \caption{Newton-Direction-Based ReLU-Thresholding (NDRT) Algorithm}
\label{RIHTwithnewtonstep}
	\begin{itemize} 
		\item  Input:   measurement matrix $\vA \in \bR^{m \times n}$, measurement vector $\vy \in \bR^m$, sparsity level $k$, parameter $\eps > 0,$ and stepsize $\lambda>0$.
        \item  Initialization: $p=0, \vx^{(0)} \in \bR^n$.
		\item  Iteration:  At the current point $\vx^{(p)}$, set        
		  \begin{enumerate}[]
		  	\item $ \vu^{(p+1)} =  
		  	\vx^{(p)} + \lambda (\t{\vA}\vA + \eps \vI)^{-1} \t{\vA} (\vy-\vA \vx^{(p)}), $
		  	\item $\vx^{(p+1)} = {\cal H}_k	(\Psi(\vu^{(p+1)}) )$, where $\Psi$ is the ReLU function.
		  \end{enumerate}
          Repeat until a stopping criterion is satisfied.
                  \item  Output:  The $k$-sparse nonnegative vector $\vx^*$.
	\end{itemize}
\end{algorithm}
\begin{algorithm} \caption{Newton-Direction-Based ReLU-Thresholding Pursuit (NDRTP) Algorithm}
\label{RHTPwithnewtonstep}
	\begin{itemize} 
		\item  Input:   measurement matrix $\vA \in \bR^{m \times n}$, measurement vector $\vy \in \bR^m$, sparsity level $k$, parameter $\eps > 0,$ and stepsize $\lambda>0$.
        \item  Initialization: $p=0, \vx^{(0)} \in \bR^n$.
		\item  Iteration:  At the current point $\vx^{(p)}$, set
        \begin{flalign} \label{NNLS}
            \vu^{(p+1)} & = \vx^{(p)} + \lambda (\t{\vA}\vA + \eps \vI)^{-1} \t{\vA} (\vy-\vA \vx^{(p)}),
            \nonumber \\
            S^{(p+1)}  & = \cL_k\(\Psi(\vu^{(p+1)}) \),\ \mbox{where}\ \Psi \ \mbox{is the ReLU function},
            \nonumber \\
            \vx^{(p+1)}  & = \argmin\limits_{\vz} \left\{ \frac{1}{2} \nnn{\vy - \vA \vz}^2 : 
                                 \supp(\vz) \subseteq S^{(p+1)}, \vz \ge 0 \right\}.
         \end{flalign}
         Repeat until a stopping criterion is satisfied.
		\item  Output:  The $k$-sparse nonnegative vector $\vx^*$.
	\end{itemize}
\end{algorithm}

 The algorithms can use different stopping criterion.  For instance, we may set the maximum number of iterations allowed to perform, or check whether the  iterate  already  matches the measurement accurately enough, i.e., $\|y-Ax^{(p)}\|_2 $ is small enough. In \cite{NM}, the stepsize $\la$ is fixed as a constant, and the regularization parameter $\eps$ is taken by $\eps = \max \{\sigma_1^2 +1, \la - \sigma_m^2\}$, where $\sigma_1, \sigma_m$ are the largest and smallest singular values of the measurement matrix $\vA$, respectively. Different from \cite{NM}, the analysis in this paper does not impose any restrictive condition on the choice of $\eps > 0$ (see the next section for details).

\section{Analysis of Algorithms}

In this section, we perform a theoretical analysis   to establish the guaranteed performance results for the proposed algorithms under certain conditions. We first improve upon a fundamental lemma from \cite{NM}, originally derived under the restrictive condition $\eps > \sigma_1^2 = \nnn{\t{\vA}\vA}$, which ensures convergence of the matrix power series:
 $$\left(\vI + \frac{1}{\eps}\t{\vA}\vA \right)^{-1} = \sum_{j = 0}^{\infty} (-1)^j (\frac{1}{\eps} \t{\vA}\vA)^j.$$
This requirement on $\eps$ is limiting both theoretically and practically. The present work removes this restriction entirely through an alternative analysis based on the following spectral property: for any real symmetric matrix $\mathbf{M} \in \mathbb{R}^{n\times n}$ with eigenvalues $ \lambda_1, \dots, \lambda_n,$ and any rational function $q$ for which $q(\mathbf{M})$ is well-defined, the eigenvalues of $q(\mathbf{M}) $ are precisely $q(\lambda_1), \dots, q(\lambda_n).$
We now state the improved version of Lemma 3.2  in \cite{NM}).

\blem \label{NMengImproved}
Let $\vA \in \bR^{m \times n}$ be a measurement matrix with $m \ll n$ and $\eps>0$ be  a given parameter.
Denote the largest and smallest singular values of $\vA$ by $\sigma_1$ and $\sigma_m$, respectively. 
Given an index set $\Omega \subseteq [n]$ and vectors $\vu, \vv \in \bR^n$, 
let $\lambda$ be a positive parameter such that $\lambda \le \sigma_m^2 + \eps$. 
Then
\be \label{RIP1}
|\langle \vu, (\vI - \la (\t{\vA} \vA + \eps \vI)^{-1}\t{\vA} \vA)\vv \rangle | 
\le  \left( \delta_k + \sigma_1^2 - \frac{\la \sigma_1^2}{\sigma_1^2 + \eps} \right) \nnn{\vu} \nnn{\vv}
\ee
if $|\supp(\vu) \cup \supp(\vv)| \le k$,
and
\be \label{RIP2}
\left\| 
\left((\vI - \la (\t{\vA} \vA + \eps \vI)^{-1}\t{\vA} \vA) \vv \right)_\Omega
\right\|_2
\le \left( \delta_s + \sigma_1^2 - \frac{\la \sigma_1^2}{\sigma_1^2 + \eps} \right) \nnn{\vv}
\ee
if $|\Omega \cup \supp(\vv)| \le s$.
\elem

\bpf  Let us start by defining the  function
$$q(t) := \( 1 - \frac{\la}{t + \eps}\) t, \quad t\ge 0.$$
Since $\t{\vA} \vA + \eps \vI$ is positive definite for any $\eps > 0$,
the matrix
$q(\t{\vA} \vA) = (\vI - \la (\t{\vA} \vA + \eps \vI)^{-1}) \t{\vA}\vA$ 
 is  well defined and symmetric. It follows that
\be
\begin{split} \label{lemmaineq1}
 |\langle {\vu}, {(\vI-\la(\t{\vA}\vA+\eps \vI)^{-1} \t{\vA}\vA)\vv} \rangle|  
      &  =  |\langle {\vu}, {(\vI - \t{\vA}\vA+ q(\t{\vA}\vA))\vv} \rangle |  \\
      & \le |\langle {\vu}, {(\vI - \t{\vA}\vA)\vv} \rangle | 
        + |\langle {\vu}, { q(\t{\vA}\vA)\vv} \rangle |   \\
      & \le \, \delta_k \nnn{\vu} \nnn{\vv}+ \nnn{q(\t{\vA}\vA)}\nnn{\vu} \nnn{\vv}, 
\end{split}
\ee
where the first term of the final inequality follows from (\ref{RIC_ineq3}) with $|\supp(\vu) \cup \supp(\vv)| \le k$.

We now  examine the norm of $q(\t{\vA}\vA)$.
Let  $\sigma_1 \ge \cdots \ge \sigma_m$ be the singular values of $\vA$, 
then the eigenvalues of $\t{\vA}\vA$ are $\sigma_1^2 \ge \dots \ge \sigma_m^2$ 
with additional $(n-m)$ zero eigenvalues. 
Therefore, the matrix $q(\t{\vA}\vA)$ has eigenvalues $q(\sigma_1^2), \cdots, q(\sigma_m^2)$, along with additional $(n-m)$ zeros. 
Since $\la \le \sigma_m^2 + \eps$, we have
$$q(\sigma_m^2)=  ( 1 - \frac{\la}{\sigma_m^2 + \eps}) \sigma_m^2 = \frac{\sigma_m^2}{\sigma_m^2 + \eps} (\sigma_m^2 + \eps -\la)\ge 0.$$ 
It follows immediately from the definition of $q(t)$ that $q$ is strictly increasing on 
$[\sigma_m^2, \infty)$.
As a result, the  eigenvalues of the matrix $q(\t{\vA}\vA)$ form 
a non-increasing sequence: 
$$q(\sigma_1^2) \ge \cdots \ge q(\sigma_m^2)\ge 0 = \dots  = 0.$$ 
Note that $q(\t{\vA}\vA)$ is a real symmetric matrix. We have
$$
\nnn{q(\t{\vA} \vA)} = \max_{i\in [m]} \left\{q(\sigma_i^2)\right\} 
 = q(\sigma_1^2) 
 = \sigma_1^2 - \frac{\la \sigma_1^2}{\sigma_1^2 + \eps}.
$$
Substituting it  into (\ref{lemmaineq1}) yields the desired inequality (\ref{RIP1}).

We now show that (\ref{RIP2}) holds as well. For any index set $\Omega \subseteq [n]$ satisfying $|\Omega \cup \supp(\vv)| \le s$, define 
$$
\vu := \left( \left( \vI - \lambda (\vA^\top \vA + \eps \vI)^{-1} \vA^\top \vA \right) \vv \right)_\Omega.
$$
Then $|\supp(\vu) \cup \supp(\vv)| \le s$ since  $\supp(\vu) \subseteq \Omega$.
It follows from (\ref{RIP1}) that
\begin{align*}
  \nnn{\vu}^2 
  & = \ip {((\vI - \la (\t{\vA} \vA + \eps \vI)^{-1}\t{\vA} \vA) \vv )_\Omega} 
          {(\vI - \la (\t{\vA} \vA + \eps \vI)^{-1}\t{\vA} \vA) \vv} \\
  & = \ip {\vu}{(\vI - \la (\t{\vA} \vA + \eps \vI)^{-1}\t{\vA} \vA) \vv}\\
  & \le \left( \delta_s + \sigma_1^2 - \frac{\la \sigma_1^2}{\sigma_1^2 + \eps} \right)
    \nnn{\vu} \nnn{\vv}.
\end{align*}
Dividing both sides by  $\nnn{\vu}$ (assuming $\vu \ne \bm{0}$, otherwise the inequality (\ref{RIP2}) holds trivially), we obtain
$$\nnn{\vu}  \le 
     \left( \delta_s + \sigma_1^2 - \frac{\la \sigma_1^2}{\sigma_1^2 + \eps} \right)
     \nnn{\vv}.$$
This is exactly the inequality (\ref{RIP2}), as desired.
\epf

We now establish the main performance result for NDRT in noisy settings.
\bthm \label{convofNDRT}
Let $\vy = \vA \vx + \ve$, where $\vA \in \bR^{m\times n}$ is a measurement matrix with $m \ll n$,
$\vx \in \bR^n$ is a nonnegative $k$-sparse signal and $\ve \in \bR^m$ is a noise vector.
Let $\sigma_1, \sigma_m$ be the largest and smallest singular values of $\vA$, respectively.
Suppose that $\vA$ satisfies that 
\be \label{RICofIHT}
 \delta_{3k} + \sigma_1^2 - \sigma_m^2 < \frac{\sqrt{5} - 1}{2} \approx 0.618.
\ee
For any given $\eps>0$, let the stepsize $\la$ be taken such that
\be \label{szofIHT}
\sigma_m^2 +  \frac{\sigma_m^2}{\sigma_1^2}\eps \le \la \le \sigma_m^2 + \eps.
\ee
Then the sequence $\{ \vx^{(p)}\}$ generated by {\rm NDRT} satisfies
\be \label{errorboundsNDRT}
\nnn{\vx^{(p)}- \vx} \le \al^p \nnn{\vx^{(0)}-\vx} + \frac{\gamma}{1-\al}  \nnn{\ve},
\ee
where
\be \label{paraNDRT}
\al := \phi\left( \delta_{3k} + \sigma_1^2 - \frac{\la \sigma_1^2}{\sigma_1^2 + \eps} \right)<1, \qquad 
\gamma :=\frac{ \phi \la \sigma_1}{\sigma_m^2 + \eps}, 
\ee
with $\phi$ being the golden ratio.
In particular, if $\ve = \bf{0}$, then $\{\vx^{(p)}\}$ converges to $\vx$.
\ethm

\bpf
For notational simplicity, we denote $\vu^{(p + 1)}$ in NDRT by $\vu$. Then we have 
\begin{align*}
\vu & = \vx^{(p)} + \lambda (\t{\vA}\vA + \eps \vI)^{-1} \t{\vA} (\vy-\vA \vx^{(p)}) \nonumber \\
    & = \vx^{(p)} + \lambda (\t{\vA}\vA + \eps \vI)^{-1} \t{\vA} (\vA (\vx- \vx^{(p)}) + \ve) \nonumber \\
    & = (\vI - \la (\t{\vA}\vA + \eps \vI)^{-1} \t{\vA} \vA )(\vx^{(p)} - \vx) + \vx + \la (\t{\vA}\vA + \eps \vI)^{-1} \t{\vA} \ve. 
\end{align*}
Let 
$$\vg:= \left(\vI - \la (\t{\vA}\vA + \eps \vI)^{-1} \t{\vA} \vA \right)(\vx^{(p)} - \vx), 
\quad
 \vh:=\la (\t{\vA}\vA + \eps \vI)^{-1} \t{\vA} \ve.$$
 Then $\vu = \vg + \vx + \vh$. 
For any given $\Omega \subseteq [n]$, we have
\begin{align}\label{keyineq}
\nnn{(\vu - \vx)_\Omega}^2 
 &  =    \nnn{(\vu^+ - \vx)_\Omega + (\vu^-)_\Omega}^2 \nonumber \\
 &  =    \nnn{(\vu^+ - \vx)_\Omega}^2 +\nnn{(\vu^-)_\Omega}^2 + 2 \ip{(\vu^+ - \vx)_\Omega}{(\vu^-)_\Omega} \nonumber \\
 & \ge  \nnn{(\vu^+ - \vx)_\Omega}^2,
\end{align}
where the inequality is obtained from 
$$\ip{(\vu^+ - \vx)_\Omega}{(\vu^-)_\Omega}
=\ip{(\vu^+)_\Omega}{(\vu^-)_\Omega} - \ip{\vx_\Omega}{(\vu^-)_\Omega}
\ge 0,$$
which follows from the nonnegativity of $\vx$.  Now, let
 $$\Omega:=\supp(\vx^{(p+1)}) \cup \supp(\vx) = \supp(\mathcal{H}_k(\vu^+)) \cup \supp(\vx).$$
 Then, we have
\be
\label{normineq4}
 \nnn{\vx^{(p+1)} - \vx} 
     =   \nnn{\cH_k(\vu^+) - \vx}
       \le \phi \nnn{(\vu^+ - \vx)_\Omega} 
    \le \phi \nnn{(\vu - \vx)_\Omega}  =  \phi \nnn{\left( \vg + \vh \right)_\Omega},
\ee
where the first and second inequalities follow from (\ref{tightbound}) and (\ref{keyineq}), respectively.
Note that $\la \le \sigma_m^2 + \eps$ and
$$|\Omega \cup \supp(\vx^{(p)} - \vx)| \le |\supp(\vx^{(p+1)}) \cup \supp(\vx) \cup \supp(\vx^{(p)})|\le 3k.$$ 
By (\ref{RIP2}), we have
\be \label{u_norm}
\nnn{\vg_{_\Omega}} 
= \left\|
 \left( \left(\vI - \la (\t{\vA}\vA + \eps \vI)^{-1} \t{\vA} \vA \right)(\vx^{(p)} 
- \vx)\right)_\Omega 
\right\|_2 
\le \left( \delta_{3k} + \sigma_1^2 - \frac{\la \sigma_1^2}{\sigma_1^2 + \eps} \right)
 \nnn{\vx^{(p)} - \vx}.
\ee
Next we determine the upper bound of $\nnn{\vh_{\Omega}}$. To this end, we first examine the norm of $(\t{\vA}\vA + \eps \vI)^{-1} \t{\vA}$. 
Suppose $\sigma_1\ge \cdots \ge \sigma_m$ are all singular values of $\vA$. Then $\t{\vA}\vA$ has eigenvalues
$\sigma_1^2, \cdots, \sigma_m^2$ along with additional $(n-m)$ zeros. Define a function 
$$Q(t) := \frac{t}{(t + \eps)^2}, \quad t\ge 0.$$
Then the matrix $((\t{\vA}\vA + \eps \vI)^{-1} \t{\vA})\t{ ((\t{\vA}\vA + \eps \vI)^{-1} \t{\vA})}= Q(\t{\vA} \vA)$ has eigenvalues $$Q(\sigma_i^2) = \frac{\sigma_i^2}{(\sigma_i^2 + \eps)^2}, \quad i=1, \cdots, m,$$
with additional $(n-m)$ zeros. Thus, we obtain
$$\nnn{(\t{\vA}\vA + \eps \vI)^{-1} \t{\vA}} = \max_{i \in [m]}  \sqrt{Q(\sigma_i^2)} 
= \max_{i \in [m]} \frac{\sigma_i}{\sigma_i^2 + \eps} 
\le \frac{\sigma_1}{\sigma_m^2 + \eps},$$
where the final inequality is ensured by $\sigma_1 \ge \cdots \ge \sigma_m$. It follows that
\be \label{w_norm}
\begin{split}
\nnn{\vh_{_{\Omega}}}
  & \le \left\|\vh\right\|_2 
     = \la \nnn{ (\t{\vA}\vA + \eps \vI)^{-1} \t{\vA} \ve} \\
  &  \le \la \nnn{(\t{\vA}\vA + \eps \vI)^{-1} \t{\vA}} \, \nnn{\ve}\\
  & \le \frac{\la \sigma_1}{\sigma_m^2 + \eps} \, \nnn{\ve}.  
\end{split}
\ee
Merging (\ref{normineq4}), (\ref{u_norm}) and (\ref{w_norm}) together, we obtain
\be \label{iter_NDRT}
\nnn{\vx^{(p+1)} - \vx}   
\le \phi  \left( \nnn{\vg_{_{\Omega}}} + \nnn{\vh_{_{\Omega}}}
    \right) \\
\le  \al \nnn{\vx^{(p)} - \vx} +
     \gamma  \nnn{\ve},
\ee
where $\alpha$ and $\gamma$ are  specified in  (\ref{paraNDRT}).

Under the conditions of the theorem, we now show that $0 \le \al <1$, i.e.,
$$
0 \le \phi  \left( \delta_{3k} + \sigma_1^2 - \frac{\la \sigma_1^2}{\eps + \sigma_1^2} \right) < 1.
$$
The nonnegativity $\al \ge 0$ follows directly from $\la \le \sigma_m^2 + \eps$ and $\sigma_1 \ge \sigma_m$.
To ensure $\al < 1$,  it suffices to require 
$$\la >\frac{\left(\delta_{3k}+\sigma_1^2-\frac{1}{\phi} \right)(\sigma_1^2 + \eps)}
{\sigma_1^2}.$$
This is guaranteed by conditions (\ref{szofIHT}) and (\ref{RICofIHT}):
$$\la \ge \sigma_m^2 + \frac{\sigma_m^2}{\sigma_1^2}\eps 
=  \frac{\sigma_m^2(\sigma_1^2 + \eps)}{\sigma_1^2}
> \frac{\left(\delta_{3k}+\sigma_1^2- \frac{1}{\phi} \right)(\sigma_1^2 + \eps)}
{\sigma_1^2}.$$
Having established that $0 \le \al <1$, we may recursively apply the inequality (\ref{iter_NDRT}) to obtain
\begin{align} \label{recursion}
\nnn{\vx^{(p)} - \vx} 
& \le  \al \nnn{\vx^{(p-1)} - \vx} +
     \gamma  \nnn{\ve} \le \cdots \nonumber \\
& \le \al^p \nnn{\vx^{(0)}-\vx} + \frac{1-\al^p}{1-\al} \gamma\nnn{\ve}\\
& \le \al^p \nnn{\vx^{(0)}-\vx} + \frac{\gamma}{1-\al} \nnn{\ve}. \nonumber
\end{align}
Clearly, $\{\vx^{(p)}\}$ converges to $\vx$ if $\ve = \bm{0}$.
\epf

\begin{rk}
 Our analysis and main result established above is significantly different from those in \cite{NM}.  The convergence result of  NSIHT in \cite{NM} (Theorem 3.4 therein)  requires that 
 $\vA$  satisfy $\delta_{3k} < 1/\sqrt{3}$ and the parameter $\eps$ satisfy
\be \label{NMeps}
\eps  > \max \left\{\sigma_1^2, 
\(\frac{\sigma_1^2 - \sigma_m^2}{\frac{1}{\sqrt{3}} - \delta_{3k}} - 1\)\sigma_1^2\right\}.
\ee
However, as shown  in Theorem \ref{convofNDRT}, the original restriction on $\eps$ such as (\ref{NMeps})  can be entirely removed under the RIP condition (\ref{RICofIHT}). This means the parameter $\eps$ can set to be any small positive number. In this case, the term $\t{\vA}\vA+\eps \vI$ can be interpreted as a perturbation of the Hessian $\t{\vA}\vA$, and hence $(\t{\vA}\vA + \eps \vI)^{-1}\t{\vA}(\vy - \vA\vx)$ can be seen as a truly
 modified Newton direction. Similarly, our analysis here can also be used to improve the  convergence result of NSHTP in \cite{NM}, i.e., Theorem 4.3 therein.
\end{rk}


Next, we investigate the guaranteed performance of  NDRTP in noisy settings.
Before stating the main results, we first establish a fundamental property of   nonnegative projection, which  plays a critical role in the subsequent analysis.
\bthm \label{KeyLem}
Let $\vy = \vA \vx + \ve$ be noisy measurements of a nonnegative $k$-sparse signal $\vx \in \bR^n$,
where the measurement matrix $\vA \in \bR^{m \times n}$ satisfies the RIP of order $2k$, and $\ve \in \bR^m$ is
a  noise vector. 
For any index set $\Lambda \subseteq [n]$ with $|\Lambda| \le k$, denote
$$\vz^* := \argmin_{\vz}\left\{ \nnn{\vy - \vA \vz} : \supp(\vz) \subseteq \Lambda, \vz \ge \bm{0} \right\}.$$
Then 
$$\nnn{\vz^*-\vx} \le \frac{1}{\sqrt{1-\delta^2_{2k}}}
                       \nnn{(\vz^*-\vx)_{\ol{\Lambda}}}
    +\frac{\sqrt{1+\delta_k}}{1-\delta_{2k}}\nnn{\ve}. $$
\ethm

\bpf Note that the constraint of $\vz^*$ can be written equivalently as
$z_i \ge 0$ for all $i \in \Lambda$ and $z_j = 0$ for all $j\in \ol{\Lambda}$.
So, $\vz^*$  is the optimal solution to the constrained minimization problem:
\begin{align*}
  \min        & \quad f(\vz):= \frac{1}{2} \nnn{\vy - \vA\vz}^2 \\
  \text{s.t.} & \quad z_i \ge 0 ~\text{for all}~ i\in \Lambda, \\
              & \quad z_j = 0 ~\text{for all}~ j\in \ol{\Lambda}.
\end{align*} 
The corresponding Lagrangian function is given by
$$
\cL(\vz, \bm{\mu}):= f(\vz) - \sum_{i \in \Lambda} \mu_i z_i - \sum_{j \in \ol{\Lambda}} \mu_j z_j,
$$
where $\bm{\mu}= \t{(\mu_1, \dots, \mu_n)} \in \bR^n$ is the vector of Lagrange multipliers.
According to the Karush-Kuhn-Tucker  optimality conditions, there exists a vector 
$\bm{\mu}^* = \t{(\mu^*_1, \cdots, \mu^*_n)} \in \bR^n$ such that
\begin{equation}\label{KKT}
\left\{
\begin{aligned}
&(\nabla f(\vz^*))_i - \mu^*_i = 0,  && i \in \Lambda, \\
&(\nabla f(\vz^*))_j - \mu^*_j = 0,  && j \in \overline{\Lambda}, \\
&z^*_i \ge 0,                            && i \in \Lambda, \\
&z^*_j = 0,    && j \in \ol{\Lambda},\\
&\mu^*_i \ge  0,\ \mu^*_i z_i^* = 0, && i \in {\Lambda}.
\end{aligned}
\right.
\end{equation}

 Let $\Lambda_1 := \supp(\vz^*) \subseteq \Lambda$ and $\Lambda_2: =\Lambda \setminus \Lambda_1$. 
By the complementary slackness condition in (\ref{KKT}), we have $\mu_i^* = 0$ for all $i \in \Lambda_1$, i.e.,
$(\bm{\mu}^*)_{\Lambda_1} = \bm{0}$. 
Combining the stationarity condition and dual feasibility in (\ref{KKT}) leads to
$(\nabla f(\vz^*))_{\Lambda} = (\bm{\mu}^*)_{\Lambda} \ge \bm{0}$. Therefore, we have
$$
(\nabla (f(\vz^*)))_{\Lambda_1} = (\bm{\mu}^*)_{\Lambda_1} = \mathbf{0}
\quad \text{and} \quad
(\nabla (f(\vz^*)))_{\Lambda_2} = (\bm{\mu}^*)_{\Lambda_2} \ge \mathbf{0}.
$$
It follows that
$$
\nnn{(\vx-\vz^*)_{\Lambda_1}}   =    \nnn{\(\vx -\vz^* + \nabla f(\vz^*)\)_{\Lambda_1}},$$
$$
\nnn{(\vx-\vz^*)_{\Lambda_2}}   \le  \nnn{\(\vx -\vz^* + \nabla f(\vz^*)\)_{\Lambda_2}},
$$
where the inequality holds due to $(\vz^*)_{\Lambda_2} = \bm{0}$ and
$\vx \ge \bm{0}$.
Thus,
\begin{align*}
\nnn{\(\vx - \vz^*\)_\Lambda}^2 
& =  \nnn{\(\vx - \vz^*\)_{\Lambda_1}}^2 + \nnn{\(\vx - \vz^*\)_{\Lambda_2}}^2\\
& \le  \nnn{\(\vx - \vz^* + \nabla f(\vz^*)\)_{\Lambda_1}}^2 
     +  \nnn{\(\vx - \vz^* + \nabla f(\vz^*)\)_{\Lambda_2}}^2 \\
&  = \nnn{\(\vx - \vz^* + \nabla f(\vz^*)\)_{\Lambda}}^2.
\end{align*}
As $\nabla f(\vz^*) = \t{\vA}(\vA \vz^*-\vy)$, where $\vy = \vA \vx + \ve$, the inequality above  becomes
\begin{align*}
& \nnn{\(\vx - \vz^*\)_{\Lambda}}^2 \\
& \le  \nnn{(\vx -\vz^* + \t{\vA}(\vA \vz^*-\vy))_{\Lambda}}^2 \\
& = \nnn{((\vI - \t{\vA}\vA)(\vx - \vz^*)-\t{\vA}\ve)_{\Lambda} }^2 \\
& = \nnn{((\vI - \t{\vA}\vA)(\vx - \vz^*))_{\Lambda} }^2 + \nnn{(\t{\vA} \ve)_{\Lambda}}^2 
    - 2 \ip{((\vI-\t{\vA}\vA)(\vx - \vz^*))_{\Lambda}}{(\t{\vA}\ve)_{\Lambda}} \\
& \le \nnn{((\vI - \t{\vA}\vA)(\vx - \vz^*))_{\Lambda} }^2 + \nnn{(\t{\vA} \ve)_{\Lambda}}^2 
    + 2 \nnn{((\vI - \t{\vA}\vA)(\vx - \vz^*))_{\Lambda}}  \, \nnn{(\t{\vA} \ve)_{\Lambda}} \\
& \le \delta_{2k}^2 \nnn{\vx - \vz^*}^2 + \nnn{(\t{\vA}\ve)_\Lambda}^2
+ 2\delta_{2k}  \nnn{\vx - \vz^*}\, \nnn{(\t{\vA}\ve)_\Lambda},  
\end{align*}
where the final inequality is obtained from (\ref{RIC_ineq2}) with $|\supp(\vx - \vz^*) \cup \Lambda| \le 2k$. Then we have
\begin{align*}
  \nnn{\vx - \vz^*}^2   
   & =  \nnn{\(\vx - \vz^*\)_\Lambda}^2 + \nnn{\(\vx - \vz^*\)_{\ol{\Lambda}}}^2 \\
   & \le \delta_{2k}^2 \nnn{\vx - \vz^*}^2 + \nnn{(\t{\vA}\ve)_\Lambda}^2 
    + 2\delta_{2k}  \nnn{\vx - \vz^*} \nnn{(\t{\vA}\ve)_{\Lambda}} + \nnn{\(\vx - \vz^*\)_{\ol{\Lambda}}}^2 ,
\end{align*}  
which implies that
$$(1-\delta_{2k}^2) \nnn{\vx - \vz^*}^2 - 2 \delta_{2k} \nnn{(\t{\vA}\ve)_\Lambda} \nnn{\vx - \vz^*} 
- \nnn{(\t{\vA}\ve)_\Lambda}^2 -
\nnn{\(\vx - \vz^*\)_{\ol{\Lambda}}}^2 \le 0.$$
The above inequality amounts to $P(\nnn{\vx - \vz^*}) \le 0$, where $P(t)$ is the quadratic function  in variable $t$ defined by
$$P(t) := (1-\delta_{2k}^2) t^2 
- 2 \delta_{2k} \nnn{(\t{\vA}\ve)_\Lambda} t 
- \nnn{(\t{\vA}\ve)_\Lambda}^2 
-\nnn{\(\vx - \vz^*\)_{\ol{\Lambda}}}^2 . $$
This implies that $\nnn{\vx - \vz^*}$ is less than the largest root of $P(t)$, i.e.,
\begin{align*}
  \nnn{\vx - \vz^*} & \le \frac{ \delta_{2k} \nnn{(\t{\vA}\ve)_\Lambda} 
  + \sqrt{\nnn{(\t{\vA}\ve)_\Lambda}^2 
  + (1-\delta_{2k}^2) \nnn{\(\vx - \vz^*\)_{\ol{\Lambda}}}^2}}{1-\delta_{2k}^2} \\
   & \le \frac{(1 + \delta_{2k})\nnn{(\t{\vA}\ve)_\Lambda}
   + \sqrt{1-\delta_{2k}^2} 
   \nnn{\(\vx - \vz^*\)_{\ol{\Lambda}}}
   }{1-\delta_{2k}^2}. \\
    & \le \frac{1}{\sqrt{1-\delta^2_{2k}}}
                       \nnn{(\vz^*-\vx)_{\ol{\Lambda}}}
    +\frac{\sqrt{1+\delta_k}}{1-\delta_{2k}}\nnn{\ve},   
\end{align*}
where the final inequality follows from (\ref{RIC_ineq1}) with $|\Lambda| \le k$,  we thereby complete the proof.
\epf

Based on Theorems \ref{convofNDRT} and \ref{KeyLem}, we now further analyze the guaranteed performance of NDRTP. The proof of the main result here is inspired by the approach in \cite{SF}.
\bthm \label{convofRNHTP}  
Let $\vy = \vA \vx + \ve$, where $\vA \in \bR^{m\times n}$ is the measurement matrix,
$\vx \in \bR^n$ is a nonnegative $k$-sparse signal and $\ve \in \bR^m$ is a noise vector.
Let $\sigma_1$ and $\sigma_m$ be the largest and smallest singular values of the matrix $\vA$. 
Suppose  $\vA$ satisfies that
\be \label{RipNDRTP}
\delta_{3k} + \sigma_1^2 - \sigma_m^2 < \frac{1}{\sqrt{3}} \approx 0.577.
\ee
Let $\eps$ be a positive number and let the stepsize $\la$ be given by (\ref{szofIHT}), i.e.,
$$
 \sigma_m^2 + \frac{\sigma_m^2}{\sigma_1^2} \eps 
 \le  \la 
 \le  \sigma_m^2 + \eps.
$$
Then the sequence $\left\{\vx^{(p)}\right\}$ generated by {\rm NDRTP} satisfies
\be \label{cvgHTP}
\nnn{\vx^{(p)}-\vx} \le \rho^p \nnn{\vx^{(0)}-\vx} +  \frac{\tau}{1-\rho}\nnn{\ve},
\ee
where
$$\rho = 
\sqrt{\frac{2}{1-\delta_{2k}^2}} \(\delta_{3k}+\sigma_1^2-\frac{\la \sigma_1^2}{\sigma_1^2 + \eps}\)
<1
$$
and
$$\tau =  \sqrt{\frac{2}{1-\delta_{2k}^2}}\, \frac{\la \sigma_1}{\sigma_m^2 + \eps} 
         + \frac{\sqrt{1 + \delta_{k}}}{1 - \delta_{2k}}.
$$
In particular,  if $\ve = \bf{0}$, then the sequence $\{\vx^{(p)}\}$ converges to $\vx$.
\ethm 

\bpf Let $\vu$ denote the positive part of $\vu^{(p+1)}$ in NDRTP, i.e., $\vu:= \Psi(\vu^{(p+1)}) = (\vu^{(p+1)})^+$.
Then $S^{(p+1)} = \cL_k(\vu)$.  Thus,
\be \label{normineq}
\begin{split}
\nnn{\vu_{S\setminus S^{(p+1)}}} 
& = \nnn{\vu_{S \setminus (S \cap S^{(p+1)})}}
 \le \nnn{\vu_{S^{(p+1)} \setminus (S \cap S^{(p+1)})}}\\
& = \nnn{\vu_{S^{(p+1)}\setminus S}}
  = \nnn{(\vu - \vx)_{S^{(p+1)}\setminus S}},
\end{split}
\ee
where $S = \supp (\vx)$.
Since $\supp (\vx^{(p+1)}) \subseteq S^{(p+1)}$ in NDRTP, we obtain
\be \label{normineq2-1}
  \nnn{(\vx^{(p+1)} - \vx)_{\ol{S^{(p+1)}}}} 
    =  \nnn{ \vx_{S \setminus S^{(p+1)}}}  
    \le \nnn{\vu_{S \setminus S^{(p+1)}}} + \nnn{(\vu - \vx)_{S \setminus S^{(p+1)}}}.
\ee
Combining (\ref{normineq}) with (\ref{normineq2-1}), we have
\be \label{normineq2-2}
\begin{split}
\nnn{(\vx^{(p+1)} - \vx)_{\ol{S^{(p+1)}}}} 
   & \le \nnn{(\vu - \vx)_{S^{(p+1)} \setminus S}} 
   + \nnn{(\vu - \vx)_{S \setminus S^{(p+1)}}} \\
   & \le \sqrt{2} \cdot \sqrt{\nnn{(\vu - \vx)_{S^{(p+1)} \setminus S}}^2
          + \nnn{(\vu - \vx)_{S \setminus S^{(p+1)}}}^2} \\
   & =\sqrt{2} \nnn{(\vu - \vx)_{\Omega}},   
\end{split}
\ee
where $\Omega := (S \setminus S^{(p+1)}) \cup (S^{(p+1)} \setminus S)$.
Evidently, the inequality (\ref{keyineq}) remains valid under our current definitions of
$\vu$, $\vx$ and $\Omega$; that is,
$$\nnn{(\vu - \vx)_{\Omega}} = \nnn{((\vu^{(p+1)})^+ - \vx)_\Omega} \le \nnn{(\vu^{(p+1)} - \vx)_{\Omega}}.$$
With $\vg$ and $\vh$ defined in the proof of Theorem \ref{convofNDRT}, we have
$\vu^{(p+1)} = \vg + \vx + \vh$ and
\be \label{normineq3}
  \nnn{(\vu - \vx)_\Omega} 
  \le \nnn{\vg_\Omega} + \nnn{\vh_\Omega}  
 \le \left( \delta_{3k} + \sigma_1^2 - \frac{\la \sigma_1^2}{\eps + \sigma_1^2} \right) \nnn{\vx^{(p)} - \vx}
      + \frac{\la \sigma_1}{\eps + \sigma_m^2}  \nnn{\ve},
\ee
where the final inequality follows from (\ref{u_norm}) and (\ref{w_norm}).
By substituting  $\vz^* = \vx^{(p+1)}$ and $\Lambda = S^{(p+1)}$ into Theorem \ref{KeyLem}, we have
\begin{align*} 
  \nnn{\vx^{(p+1)} - \vx}
   & \le \frac{1}{\sqrt{1 - \delta_{2k}^2}} \nnn{(\vx^{(p+1)} - \vx)_{\ol{S^{(p+1)}}}}  
     + \frac{\sqrt{1+ \delta_k}}{1-\delta_{2k}} \nnn{\ve}.
\end{align*}
Combining the above result with 
(\ref{normineq2-2}) and (\ref{normineq3}) yields
\be
\begin{split} \label{iter_NDRTP}
  \nnn{\vx^{(p+1)} - \vx}
   & \le \sqrt{\frac{2}{1 - \delta_{2k}^2}} \nnn{(\vu - \vx)_{\Omega}} 
    + \frac{\sqrt{1+\delta_k}}{1- \delta_{2k}} \nnn{\ve} \\
   & \le \rho \nnn{\vx^{(p)} - \vx} + \tau \nnn{\ve}, 
\end{split}
\ee
where $\rho$ and $\tau$ are given exactly as in Theorem \ref{convofRNHTP}. 
The condition $\la \le \sigma_m^2 + \eps \le \sigma_1^2 + \eps$ implies  $\rho \ge 0$. 
Applying an analysis similar to that in \eqref{recursion}, we obtain
$$\nnn{\vx^{(p)} - \vx} 
\le \rho^p \nnn{\vx^{(0)}-\vx} + \frac{1-\rho^p}{1-\rho} \tau \nnn{\ve}
\le \rho^p \nnn{\vx^{(0)}-\vx} + \frac{\tau}{1-\rho} \nnn{\ve}.
$$
 We now  prove that
$$\rho = 
\sqrt{\frac{2}{1-\delta_{2k}^2}} \(\delta_{3k}+\sigma_1^2-\frac{\la \sigma_1^2}{\varepsilon+\sigma_1^2}\)
<1.$$
By (\ref{RipNDRTP}) and $\sigma_1 \ge \sigma_m$, we have $\delta_{3k} < 1/\sqrt{3}$.  This together with the monotonicity property $\delta_{2k} \le \delta_{3k}$ implies that
$$\sqrt{\frac{2}{1-\delta_{2k}^2}} \le \sqrt{\frac{2}{1-\delta_{3k}^2}} < \sqrt{\frac{2}{1-(1/\sqrt{3})^2}} = \sqrt{3}.$$
To ensure $\rho < 1$, it is sufficient to show that
$$\delta_{3k}+\sigma_1^2-\frac{\la \sigma_1^2}{\varepsilon+\sigma_1^2} < \frac{1}{\sqrt{3}},$$
which amounts to  
$$\la > \frac{(\delta_{3k} + \sigma_1^2 - \frac{1}{\sqrt{3}}) (\eps + \sigma_1^2)}{\sigma_1^2}.$$
This  is guaranteed by (\ref{RipNDRTP}) together with the lower bound of $\la$ in (\ref{szofIHT}). In fact,
$$\la \ge \frac{\sigma_m^2}{\sigma_1^2}\eps + \sigma_m^2 
=  \frac{\sigma_m^2(\eps + \sigma_1^2)}{\sigma_1^2}
> \frac{\left(\delta_{3k}+\sigma_1^2-\frac{1}{\sqrt{3}}\right)(\eps+\sigma_1^2)}
{\sigma_1^2}.$$
This completes the proof.
\epf

\section{Numerical Experiments}

In this section, we provide  numerical results to demonstrate the performance of the proposed algorithms in recovering nonnegative sparse signals. We compare NDRT and NDRTP against four existing algorithms—RHT~\cite{HSWW}, RHTP~\cite{PSRP}, NNOMP~\cite{AEM}, and NNSP~\cite{KH}—using success frequency and CPU time.
All  experiments were conducted in MATLAB(R2024a) on a computer equipped with an Intel(R) Core(TM) i7-9700 processor (3.0 GHz) and 16 GB of RAM.  

Our experiments use measurement matrices $\vA \in \mathbb{R}^{m \times n}$ with entries independently drawn from $\cN(0, 1/m)$,  where $m = 600$ and $n = 2000$.
The target signal $\vx^* \in \mathbb{R}^n$ is a nonnegative $k$-sparse vector, whose nonzero entries are the absolute values of independent  standard normal random variables, with their positions chosen uniformly at random.
 The algorithm performance is evaluated under both noiseless and noisy conditions. 
The measurements are given by $\vy = \vA \vx^*$ and $\vy = \vA \vx^* + 10^{-4} \vh$, respectively, where $\vh$ is a normalized standard Gaussian random vector. All algorithms are initialized with $\vx^{(0)} = \bm{0}$.  In our experiments, 
a signal is recovered  if the relative error satisfies that
$$\frac{\nnn{\vx^{(p)} - \vx^*}}{\nnn{\vx^*}} \le 10^{-4},$$ 
where $\vx^{(p)}$ denotes the  solution produced by the algorithm  and $\vx^*$ is the true signal. 
For each sparsity level $k = 5t$ with $t = 1, \dots, 80$, we generate 50 random trials of $(\vA, \vx^*)$ or $(\vA, \vx^*, \vh)$  to estimate  success frequencies and average runtime of the algorithms. 

\subsection{Gradient Projection Algorithm for Nonnegative Least Squares Subproblems}

The NNOMP, NNSP, and NDRTP  involve solving the NNLS subproblem (\ref{NNLS}), which
can be reformulated as:
\begin{equation}
\label{nnls}
\vw^* = \argmin_{\vw \in \bR^k} \left\{\frac{1}{2} \|\vy - \vB\vw\|_2^2 \ : \ \vw \ge \bm{0} \right\},
\end{equation}
where $\vB \in \bR^{m \times k}$. 
We employ a gradient projection approach to solve such a NNLS problem. The details are outlined as follows.
\begin{algorithm} [H]
\caption*{{\bf Gradient Projection Algorithm} for nonnegative least squares problem}
	\begin{itemize} \label{grdproj}
		\item  Input:  sparsity level $k$, matrix $\vB \in \bR^{m \times k}$, measurement vector $\vy \in \bR^m$.
        \item  Initialization: $j=0, \vw^{(0)}=\bm{0}\in \bR^k, C = 0.6$, Maxiters $= 300$, $\eta_1 = 10^{-6}, \eta_2 = 10^{-8}$.
		\item  Iteration: At each iteration $\vw^{(j)}$, set
		\begin{enumerate}[]
			\item  $\va^{(j)} =  \t{\vB} (\vy-\vB \vw^{(j)})$,
            \item  $d^{(j)}_i = 
            \begin{cases}
              0, &   \text{if~} a^{(j)}_i <0 \text{~and~} \left|w^{(j)}_i\right| < \eta_1.\\
              a^{(j)}_i,  & \text{Otherwise.}
            \end{cases}$
            \item $\beta = \min\left\{C, -\max\{ \frac{w^{(j)}_i}{d^{(j)}_i} \text{~for all~} i \text{~such that~} d^{(j)}_i<0 \}\right\}$.
			\item $\vw^{(j+1)} = \vw^{(j)} + \beta \vd^{(j)}$.
            \item $j = j + 1$.
		\end{enumerate}
        Until $j =$ Maxiters  or $\nnn{\vw^{(j+1)} - \vw^{(j)}}<\eta_2$.
		\item  Output:  The nonnegative  vector $\vw^*$.
	\end{itemize}
\end{algorithm}

In practical implementations, we initialize $\vB = \vA_{S^{(p+1)}}$, 
where $\vA_{S^{(p+1)}}$ denotes the submatrix obtained by retaining the columns of $\vA$ indexed by 
the set $S^{(p+1)}$, which is updated in each iteration.
Suppose that $S^{(p+1)} = \{i_1, \dots, i_k\}$ with $1\le i_1 < \cdots < i_k \le n$.
After solving the NNLS problem (\ref{nnls}), the entries of $\vx^{(p+1)}$ in (\ref{NNLS}) can be
calculated as
$$(\vx^{(p+1)})_{i_j} = (\vw^*)_j \quad \text{for } j = 1, \dots, k;
\qquad
(\vx^{(p+1)})_i = 0 \quad \text{for } i \in \overline{S^{(p+1)}}.$$
This yields a $k$-sparse nonnegative vector $\vx^{(p+1)}$ and completes one iteration of the algorithm.

\subsection{Parameter configurations of NDRT and NDRTP}

NDRT  is observed to be highly sensitive to the stepsize $\la$ and the regularization parameter $\eps$.
Experiment results show that $\eps = 0.1$ and $\la \in [2, 3]$ yield relatively strong empirical performance for NDRT. 
Thus, we adopt $(\la, \eps) = (2, 0.1)$ as a default choice for NDRT in the subsequent experiments.

According to Theorem \ref{convofRNHTP}, determining the stepsize $\la$ of NDRTP
requires computing the largest and smallest singular values of the measurement matrix  $\vA$, which increases the computational burden. Based on this theorem, setting the stepsize to the square of the largest singular value is a  reasonable choice. However,  according to random matrix theory \cite{RV},  for an $m\times n$  Gaussian matrix with entries drawn from $\cN(0, 1/m)$, the largest singular value for such a matrix can be roughly estimated as $\sigma_1 \approx 1 + \sqrt{n/m}$.  Thus, to reduce the computational cost, we may simply set the stepsize in NDRTP as 
\be 
\label{empstepsize}
\la = \left\lceil \(1 + \sqrt{\frac{n}{m}}\)^2 \right\rceil.
\ee
 Numerical experiments demonstrate that such an empirical formula for stepsize indeed yields satisfactory performance of our algorithms.

\begin{figure}[htbp]
    \centering
    \begin{subfigure}[b]{0.48\textwidth}
        \includegraphics[width=\textwidth]{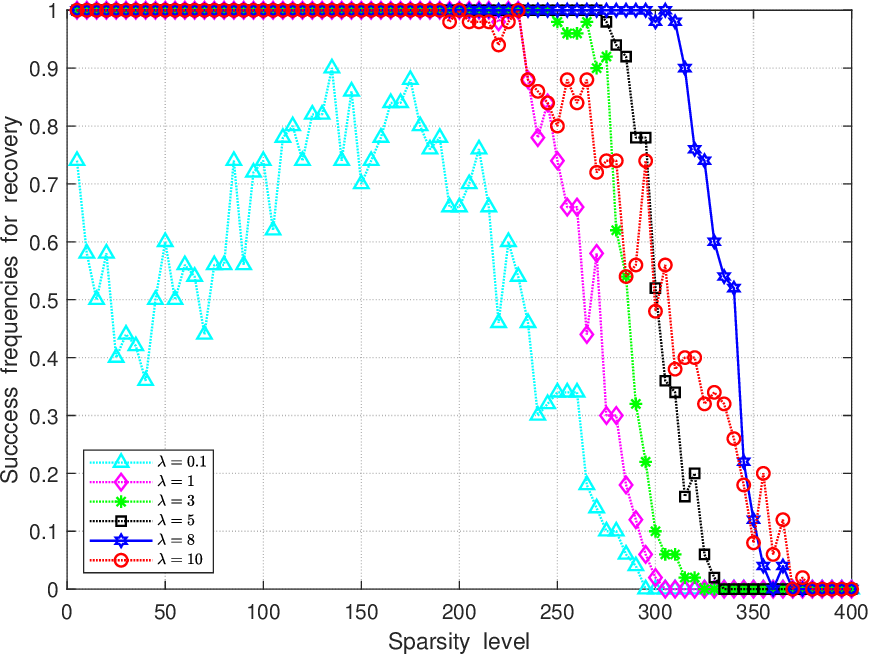}
        \caption{Algorithm performance with varying stepsizes when $\eps = 0.5$}
        \label{fig: suc_tune}
    \end{subfigure}
    \hfill
    \begin{subfigure}[b]{0.48\textwidth}
        \includegraphics[width=\textwidth]{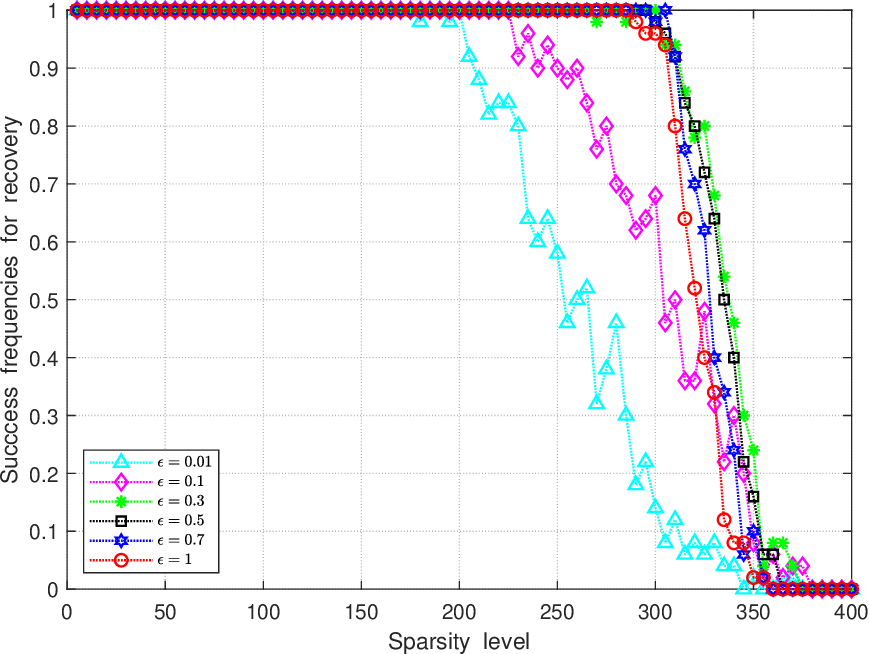} 
        \caption{Algorithm performance with varying $\eps$ when $\la = 8$}
        \label{fig: Suc_tune_eps}
    \end{subfigure}
    \caption{Success frequencies of NDRTP with varying parameters  in noiseless settings}
    \label{fig:1}
\end{figure}

To illustrate how the stepsize $\la$ might  affect the recovery performance of NDRTP, we test the algorithms using  noiseless measurements  $\vy = \vA \vx^*$ and different values  $\lambda = 0.1, 1, 3 ,5, 8, 10$ and  fixed $\eps = 0.5$. 
The  results in Fig.~\ref{fig:1}(a) indicate that the algorithm's recovery capability improves as $\la$ increases from $0.1$ to $8$, but deteriorates when $\la$ increases further to 10.  Indeed, substituting 
$m = 600$ and $n=2000$ into the empirical formula (\ref{empstepsize}) yields 
$\la = 8$, which matches the observed optimum, validating the  effectiveness of this empirical formula. 
 
To evaluate the effect of parameter $\eps$ on the  recovery performance of NDRTP, we conduct experiments with $\la = 8$ (as derived from (\ref{empstepsize})) and $\eps = 0.01, 0.1, 0.3, 0.5, 0.7, 1$.
The results are presented in Fig.~\ref{fig:1}(b).
These results demonstrate that the recovery capability of the algorithm is significantly enhanced when $\eps$ is increased from $0.01$ to $0.3$, and remains comparable when $\eps$ is between $0.3$ and $1$. 

For the remaining experiments, the  parameters of NDRTP are taken as $(\la, \eps) = (8, 0.5)$.

\subsection{Comparison to some existing algorithms}

We also carried out experiments to compare the performance of the proposed algorithms with several existing algorithms using accurate and inaccurate measurements respectively.
The parameters of  algorithms are taken as follows.
For the choice of stepsize, we set   $\la = 0.6 - \frac{k}{2m}$  for RHT as  specified in \cite{HSWW} and  $\la = 1.6$ for RHTP as  in \cite{PSRP}.
For RHT, NNSP, and NDRT, the maximum number of iterations was set to $m$; for NNOMP, it was set to the sparsity level $k$; and for RHTP and NDRTP, it was set to 50.
For  each given sparsity level,  the average CPU time was taken on successful trials only, and then  we apply a base-2 logarithmic transformation to enhance the visualization of timing differences between  algorithms.

\begin{figure}[htbp]
    \centering
    \begin{subfigure}[b]{0.48\textwidth}
        \includegraphics[width=\textwidth]{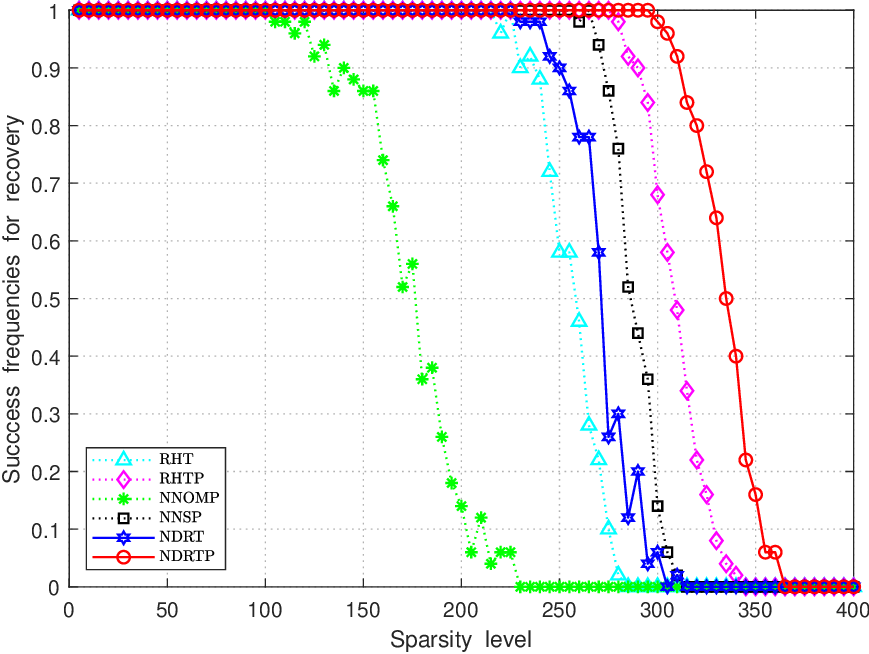}
        \caption{Success frequencies}
        \label{fig: suc_noless}
    \end{subfigure}
    \hfill
    \begin{subfigure}[b]{0.48\textwidth}
        \includegraphics[width=\textwidth]{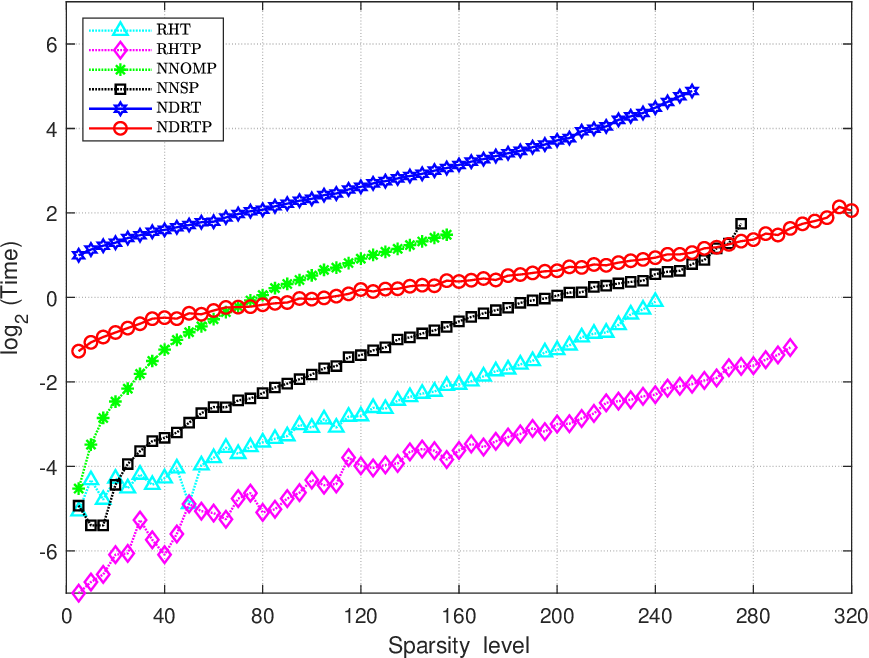}
        \caption{Average CPU time on a logarithmic scale}
        \label{fig: time_noless}
    \end{subfigure}
    \caption{Comparison of success frequencies and runtime of algorithms in noiseless settings.}
    \label{fig:2}
\end{figure}

The success frequencies for  recovery using accurate measurements are shown in Fig.~\ref{fig:2}(a).
When the sparsity level $k = 310$, the success frequency of NDRTP exceeds $90\%$, while that of RHTP drops below $50\%$, and the success frequencies of other algorithms have already dropped below $10\%$. 
At sparsity level $k=250$, NDRT achieves a success frequency of $90\%$. In contrast, the success frequencies of RHT and NNOMP drop significantly to below $60\%$ and $0\%$, respectively.
Furthermore, Tab.~\ref{tab:sparsity_levels}  lists the maximum  sparsity levels at which the success frequency reaches at least $90\%$, $80\%$, and $50\%$, respectively.
\begin{table}[htbp] \label{AccurateTab}
\centering
\caption{Maximum sparsity levels at which the success frequency is at least 90\%, 80\%, and 50\%, respectively, in noiseless settings.}
\label{tab:sparsity_levels}
\begin{tabular}{|c|c|c|c|c|c|c|}
\hline
\diagbox{\tiny{Success Frequency}}{\tiny{Algorithm}} 
& RHT & RHTP & NNOMP & NNSP & NDRT & NDRTP \\
\hline
90\% & 235 & 290 & 130 & 270 & 250 & {\bf 310} \\
\hline
80\% & 240 & 295 & 155 & 275 & 255 & {\bf 320} \\
\hline
50\% & 255 & 305 & 175 & 285 & 270 & {\bf 335} \\
\hline
\end{tabular}
\end{table}

As shown in Tab.~\ref{tab:sparsity_levels}, NDRTP achieves the highest sparsity level for any given success frequency. Meanwhile, NDRT attains a sparsity level higher than that of RHT and NNOMP, but lower than that of the remaining algorithms. These results demonstrate that NDRTP outperforms other algorithms in nonnegative sparse signal recovery. However, NDRT exhibits stronger recovery capability than RHT and NNOMP but weaker than the other algorithms.

Consistent with Fig.~\ref{fig:2}(a), Fig.~\ref{fig:2}(b) shows the runtime of each algorithm  only when the success frequency for recovery reaches at least  $80\%$. 
Consequently,  each curve terminates once its corresponding sparsity threshold is reached,  and these terminal points align with the data presented in Tab. \ref{tab:sparsity_levels}.
From Fig.~\ref{fig:2}(b) it can be observed that NDRTP requires less runtime than NDRT, but more than RHT and RHTP.
Moreover, NDRTP consumes less time than NNOMP and NNSP for relatively large sparsity levels, but this trend reverses for smaller $k$.

\begin{figure}[htbp]
    \centering
    \begin{subfigure}[b]{0.48\textwidth}
        \includegraphics[width=\textwidth]{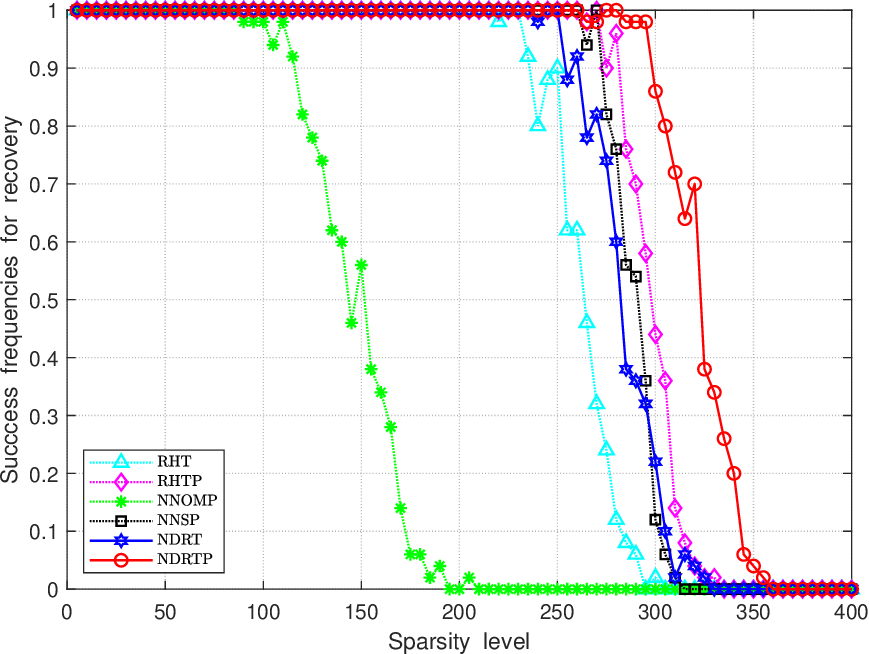}
        \caption{Success frequencies}
        \label{fig: Suc_noise}
    \end{subfigure}
    \hfill
    \begin{subfigure}[b]{0.48\textwidth}
        \includegraphics[width=\textwidth]{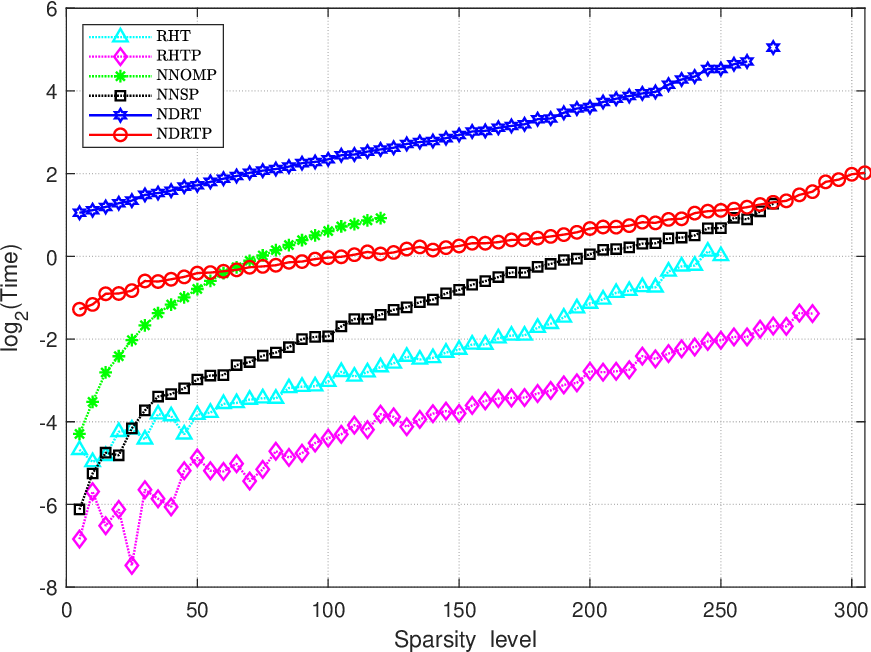}
        \caption{Average CPU time on a logarithmic scale}
        \label{fig: time_noise}
    \end{subfigure}
    \caption{Comparison of success frequencies and runtime of algorithms in noisy settings.}
    \label{fig:3}
\end{figure}

Finally, we demonstrate the  performance of algorithms using inaccurate measurements $\vy = \vA \vx^* + 10^{-4} \vh$. 
Compared to Fig.~\ref{fig:2}(a), Fig.~\ref{fig:3}(a) reveals that the noise causes a measurable degradation in the recovery performance of all algorithms. NNOMP exhibits the most severe degradation because its $k$-step iteration struggles to  capture the support of the target signal accurately under noise interference. In contrast, the other algorithms exhibits  only slight performance declines and NDRTP still performs better than the others in this experiment. 
The runtime comparison  is shown in Fig.~\ref{fig:3}(b), which is similar to that in Fig.~\ref{fig:2}(b).

\section{Conclusion}

Two novel algorithms integrating ReLU activation with Newton-type thresholding for nonnegative sparse signal recovery were proposed. The convergence of the proposed algorithms was established under joint conditions on the restricted isometry constant, measurement matrix singular values, and stepsize selection. Simulations demonstrate NDRTP achieves superior recovery performance compared to several existing methods, despite higher per-iteration computational cost. At higher sparsity levels, NDRTP also exhibits faster runtime than NDRT, NNOMP, and NNSP, and all algorithms except NNOMP remain relatively stable when the measurements are slightly inaccurate.

\end{document}